\PassOptionsToPackage{table,xcdraw}{xcolor}

\documentclass[nonacm]{acmart}

\usepackage{booktabs} 

\usepackage{mathtools}

\usepackage{amsmath}
\usepackage{algpseudocode}
\usepackage{tabularx}
\usepackage{bm}
\usepackage{cancel}
\usepackage{calc}
\usepackage{accents}
\usepackage{tikz}
\usepackage{pifont}
\usetikzlibrary{backgrounds}
\usetikzlibrary{decorations.pathreplacing, arrows.meta, tikzmark}
\usetikzlibrary{calc}
\usetikzlibrary {shapes.geometric} 
\usepackage{wrapfig}
\usetikzlibrary {fit}
\usepackage{amsmath}
\usepackage{cleveref}
\usepackage{tikzscale}
\usepackage{graphicx}
\usepackage{units} 
\usepackage[percent]{overpic}
\usepackage{subfig}
\usepackage[export]{adjustbox}
\usepackage{rotating} 

\usepackage{pgfplots}
\usepgfplotslibrary{groupplots}
\pgfplotsset{compat=1.18}

\usepackage{booktabs} 
\usepackage{makecell}

\DeclareMathOperator*{\argmax}{arg\,max}
\DeclareMathOperator*{\argmin}{arg\,min}

\tikzset{every node/.style={font=\fontfamily{LinuxBiolinumT-TLF}\selectfont}}
\usepackage{bm}
\usetikzlibrary{spy}

\newlength{\subcolumnwidth}

\newcommand{\nextsubcolumn}[1][]{%
  \cr\noalign{\hfill}
  \if\relax\detokenize{#1}\relax\else\hsize=#1\setlength{\subcolumnwidth}{\hsize}\fi
}

\usepackage[ruled]{algorithm2e} 

\SetAlFnt{\small}
\SetAlCapFnt{\small}
\SetAlCapNameFnt{\small}
\SetAlCapHSkip{0pt}

\acmJournal{TOG}

\usepackage{tikzducks}
\usepackage{duckuments}

\makeatletter
\newcommand{\showfontinfo}{
    Font: \f@family \f@series \f@shape, Size: \f@size
}
\makeatother
\begin{document}
\title{Precise Gradient Discontinuities in Neural Fields for Subspace Physics}

\author{Mengfei Liu}
\orcid{0000-0001-5879-0867}
\affiliation{%
  \institution{University of Toronto}
  \country{Canada}}
  \authornote{Indicates joint first authors.}
\email{mengfei.liu@mail.utoronto.ca}

\author{Yue Chang}
\orcid{0000-0002-2587-827X}
\affiliation{%
  \institution{University of Toronto}
  \country{Canada}}
\email{changyue.chang@mail.utoronto.ca}
\authornotemark[1]

\author{Zhecheng Wang}
\orcid{0000-0003-4989-6971}
\affiliation{%
  \institution{ University of Toronto}
  \country{Canada}}
\email{zhecheng@cs.toronto.edu}

\author{Peter Yichen Chen}
\orcid{0000-0003-1919-5437}
\affiliation{%
  \institution{MIT CSAIL}
  \country{USA}}
\email{pyc@csail.mit.edu}

\author{Eitan Grinspun}
\orcid{0000-0003-4460-7747}
\affiliation{%
  \institution{ University of Toronto}
  \country{Canada}}
\email{eitan@cs.toronto.edu}

\begin{abstract}
Discontinuities in spatial derivatives appear in a wide range of physical systems, from creased thin sheets to materials with sharp stiffness transitions. Accurately modeling these features is essential for simulation but remains challenging for traditional mesh-based methods, which require discontinuity-aligned remeshing---entangling geometry with simulation and hindering generalization across shape families.

Neural fields offer an appealing alternative by encoding basis functions as smooth, continuous functions over space, enabling simulation across varying shapes. However, their smoothness makes them poorly suited for representing gradient discontinuities. Prior work addresses discontinuities in function values, but capturing sharp changes in spatial derivatives while maintaining function continuity has received little attention.

We introduce a neural field construction that captures gradient discontinuities without baking their location into the network weights. By augmenting input coordinates with a smoothly clamped distance function in a lifting framework, we enable encoding of gradient jumps at evolving interfaces. 

This design supports discretization-agnostic simulation of parametrized shape families with heterogeneous materials and evolving creases, enabling new reduced-order capabilities such as shape morphing, interactive crease editing, and simulation of soft-rigid hybrid structures.
We further demonstrate that our method can be combined with previous lifting techniques to jointly capture both gradient and value discontinuities, supporting simultaneous cuts and creases within a unified model.
\end{abstract}

%
%


%
%


\begin{teaserfigure}
    \centering
    \includegraphics[width=1.0\linewidth]{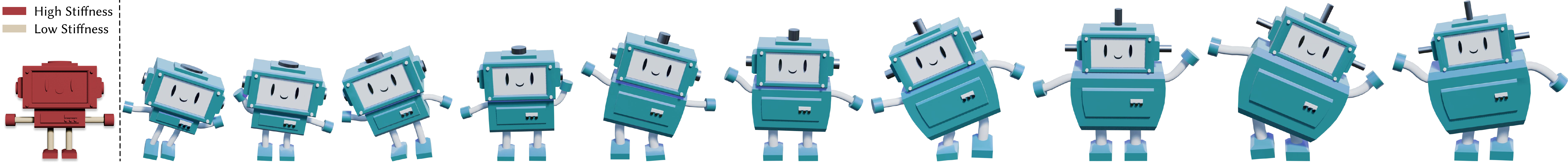}
    \caption{\emph{Dancing through life.} We introduce a neural field construction capable of representing discontinuities in spatial derivatives. Our approach allows both the domain and its internal interfaces to be parameterized over a shape space. This enables discretization-agnostic reduced-space simulation of heterogeneous materials over parametric shape families. In this animation, a stiff-bodied, soft-limbed robot dances its way from childhood to adulthood.}
    \label{fig:teaser}
\end{teaserfigure}

\maketitle

\section{Introduction}
\label{sec:introduction}

\begin{figure*}
\centering
\includegraphics[width = \linewidth]{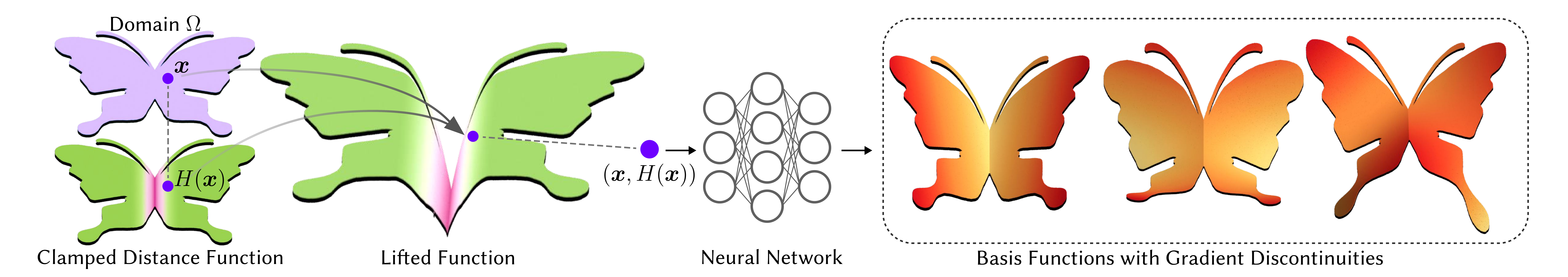}
\caption{Our method represents functions with discontinuous gradients by lifting the input domain into a higher-dimensional space. Starting from an input domain with internal interfaces, we construct a smoothly clamped distance field to augment the spatial coordinates. This defines a lifted domain where a neural network is trained to produce smooth basis functions. When restricted back to the original domain, the resulting basis captures sharp gradient transitions at the interface.}
\label{fig:overall_pipeline}
\end{figure*}

Reduced-order modeling (ROM)~\cite{barbivc2005real} accelerates physical simulation by approximating high-dimensional dynamics with a low-dimensional modal subspace. However, it remains challenging for ROM to handle discontinuities in spatial derivatives, such as those arising from folds in creased materials or interfaces within heterogeneous solids. These sharp transitions introduce localized stiffness and high-frequency behavior that global basis often fail to capture effectively.

Mesh-based simulators capture such behavior by aligning discretizations with discontinuity interfaces, but this tightly couples geometry with simulation. As the interface evolves due to shape morphing, material variation, or creasing, the mesh must be rebuilt. This invalidates earlier ROM basis and limits the model’s ability to generalize.

Several strategies attempt to decouple simulation from mesh topology. Extended finite element methods (XFEM)~\cite{Moes:1999:XFEM} enrich fixed meshes with discontinuous basis functions, allowing interfaces to move without remeshing. However, XFEM still requires a fixed background mesh and typically operates at full resolution, limiting its utility for reduced-order modeling or fast generalization across shape families.

Neural representations~\cite{Modi:2024:Simplicits, chang:2023:licrom} offer an alternative: basis functions are modeled as continuous neural fields over spatial coordinates, making them agnostic to mesh discretization. This enables subspace simulation across parametric shape families~\cite{chang2024neuralrepresentationshapedependentlaplacian}. However, their smoothness makes them ill-suited for discontinuities. While some approaches~\cite{Belhe:2023:DiscontinuityAwareNeuralFields, liu20242d} attempt to incorporate known discontinuity locations via feature alignment or differentiable fields, this embed geometric assumptions into network weights, reducing flexibility. As a result, even minor changes to the interface layout typically require retraining.

Recently, \citet{chang2025liftingwindingnumberprecise} proposed augmenting input coordinates with a generalized winding number field to model discontinuities in function values, effectively lifting the domain. This enables neural fields to handle cutting phenomena in a reusable, generalizable way. We adopt a similar lifting construction, but target a different challenge: modeling \emph{gradient discontinuities}---sharp changes in derivative, with continuity in the function itself. These arise naturally in simulations involving heterogeneous materials and creases and are not addressed by prior lifting techniques~\cite{Belhe:2023:DiscontinuityAwareNeuralFields, liu20242d, chang2025liftingwindingnumberprecise}.

The emergence of gradient discontinuities in solutions to weighted Laplace problems is well-established in the theory of elliptic PDEs: eigenfunctions of weighted Laplace operators can exhibit gradient discontinuities across interfaces where the weight function jumps~\cite{Gilbarg:2001:Elliptic}. However, this phenomenon remains underappreciated in the reduced-modeling and neural field literature. We highlight this connection and use it to motivate our construction, which explicitly encodes such discontinuities into the neural field representation.

\paragraph{Contributions}

We introduce a neural field construction for representing gradient discontinuities in a generalizable, simulation-ready form. Our key insight is to encode discontinuities via a lifting strategy that augments spatial coordinates with a non-trainable, smoothly clamped distance function. This approach enables sharp changes in function gradients without embedding interface geometry into network weights—allowing the same neural representation to be reused across families of shapes and materials.

We apply this construction to reduced-order simulation, enabling applications not previously demonstrated in this setting. Our method supports differentiable modeling of internal interfaces, allowing inverse design and real-time simulation under dynamic shape and material variation. It also enables interactive editing of creases and hybrid modeling of discontinuities in both function values and derivatives. 

In summary, we present:
\begin{itemize}
    \item a neural field architecture for encoding spatial gradient discontinuities via input lifting;
    \item a discretization-agnostic basis that generalizes across shape and material spaces;
\end{itemize}
and the first reduced-order simulation method to support
\begin{itemize}
    \item combined cuts and folds, 
    \item evolving creases, and
    \item heterogeneous materials with parameterized interfaces.
\end{itemize}





\section{Related Work}

\subsection{Discontinuity Representations}

In physical simulation, gradient discontinuities are commonly handled by aligning the mesh discretization with underlying interfaces. This is a standard approach in modeling heterogeneous materials—where the mesh conforms to stiffness boundaries \cite{Kim:2022:course}—and in simulating creases, where folding patterns are embedded into the mesh topology \cite{ZHU2021106537, Choi:2004:MSSC, Narain:2013:FCAS}. However, such mesh-dependent methods are typically limited to fixed discontinuity configurations on a single rest shape. When either the rest shape or the location of the discontinuities changes, remeshing becomes necessary. This process alters the structure of the system’s Hessian, hapering maintainance of a consistent basis and in turn reduced-order simulation.

Extended finite element methods (XFEM)~\cite{Moes:1999:XFEM, Kaufmann:2019:enrich, xfemCutting2024} avoid remeshing by enriching a fixed mesh with basis functions that capture local discontinuities, including in gradients. XFEM is typically implemented at full resolution, making it less suitable for real-time or shape-varying reduced-order simulation. Like XFEM, we also target representation of gradient discontinuities, however, we do not rely on a fixed background mesh. By fully decoupling geometry from simulation degrees of freedom, we generalize to shape families with varying rest domains (see Fig.~\ref{fig:robot}), and support reduced-order modeling. 

At the core of our method is the use of neural fields to model gradient discontinuities. Prior neural approaches, such as \citet{Belhe:2023:DiscontinuityAwareNeuralFields}, define feature fields over triangle meshes aligned to discontinuity interfaces, with subsequent work introducing differentiability \cite{liu20242d}. Rather than learning mesh-aligned features, \citet{chang2025liftingwindingnumberprecise} proposed lifting input coordinates via a generalized winding number field to model discontinuities for progressive cut simulation. These efforts primarily focus on modeling discontinuities in the function values. In contrast, our work targets a different challenge: modeling functions that remain continuous but exhibit discontinuous gradients—an essential feature for simulating heterogeneous materials and sharp features such as creases.

\subsection{Heterogeneous Elastodynamics}
Heterogeneous elastodynamics refers to the simulation of elastic materials with spatially varying properties, such as stiffness. A common strategy is to perform full-space finite element simulations, which are often computationally intensive. To improve efficiency, previous work has explored spatial simplification techniques—such as mesh coarsening \cite{chen:2017:DNC} and remeshing \cite{Chen:2018:NCD, Chen:2015:DDFE, Kharevych:2009:NCEM}—to reduce the number of degrees of freedom. Alternatively, \citet{Ty:2022:mfem} introduced a mixed discretization scheme that improves solver convergence without modifying the mesh. While these methods focus on optimizing full-space simulations of a single shape, our approach enables reduced simulation and generalizes across families of shapes.

Another line of work aims to accelerate simulations by reducing the dimensionality of the dynamical system itself. Reduced space methods \cite{barbivc2005real, Pentland:1989:GoodVibrations, benchekroun2023FastComplemDynamics, trusty:2023:subspacemfem} construct a linear subspace to eliminate redundant degrees of freedom while maintaining visually accurate motion. Our method can be seen as a generalization of reduced space simulation that extends across both shape and material spaces. \citet{Mukherjee:2016:IDSR} proposed an approach for updating the basis via incremental linear modal analysis. However, due to the matrix-based formulation, it is unclear how their method handles changes in mesh resolution or vertex count. In contrast, our model is designed to operate across varying discretizations, including different numbers of vertices.

\subsection{Neural Physics Simulation}
In recent years, neural networks have become powerful tools for accelerating physical simulation. They have been applied across a wide range of domains, including fluid modeling and reconstruction \cite{kim2019deep, Chu2022Physics, deng2023neural, Wang2024PICT, 10.1145/3641519.3657438, tao2024neural}, collision handling \cite{Romero:LCCHSD:2021, yang2020learning, Cai:2022:CSDF}, and cloth simulation \cite{Bertiche:2022:NCS, kair2024neuralclothsim, Zhang:2024:NGDSR}.

Our work is more closely related to approaches that apply neural networks to accelerate deformable simulation in reduced spaces. \citet{Holden:2019:SNP}, for example, learn nonlinear dynamics directly in a reduced coordinate system. In contrast, we preserve the physical formulation and focus on learning spatial variation. \citet{Fulton:LSD:2018} introduce a learned nonlinear basis to better capture deformation behavior, with subsequent extensions improving expressivity and generalization \cite{shen2021high, Sharp:2023:datafree, lyu2024accelerate}. While our approach uses a linear basis, it is designed to generalize across shapes and discretizations, going beyond the scope of earlier methods typically restricted to a single shape.

To support discretization-agnostic simulation, neural fields have gained popularity as spatial representations~\cite{chen2023crom, chang:2023:licrom, Modi:2024:Simplicits, chang2024neuralrepresentationshapedependentlaplacian}. Their continuous nature, however, makes it difficult to model discontinuities. The challenge of generalizing learned weights across different types of discontinuities remains largely unexplored. \citet{chang2025liftingwindingnumberprecise} address this by proposing a method to represent a family of cuts using lifted coordinates in neural fields. Like that work, our method also tackles discontinuities within continuous neural fields using input lifting. However, while they focus on function-value discontinuities (e.g., for cut simulations), we address cases where the function is continuous but its gradients are not, such as in simulations involving heterogeneous materials or sharp creases.

\section{Representing Gradient Discontinuities}

Our goal is to construct neural fields capable of accurately capturing gradient discontinuities. Let $f : \Omega \rightarrow \mathbb{R}$ be a function defined over a domain $\Omega \subset \mathbb{R}^d$, where $d$ is 2 or 3. We consider cases where the normal gradient of $f$ is discontinuous across an internal interface $\Gamma$: 
\begin{align}
\lim_{\bm{x} \to \bm{x_0}^+} \frac{\partial f(\bm{x})} {\partial \mathbf{n}} \neq \lim_{\bm{x} \to \bm{x_0}^-} \frac{\partial f(\bm{x})} {\partial \mathbf{n}} \ , \quad \bm{x}_0 \in \Gamma \ , \quad \bm{x} \in \Omega \ ,
\end{align}
with $\mathbf{n}$ denoting the normal to the interface.

\paragraph{Modeling the Gradient Discontinuity through Lifting}
We adopt a similar lifting approach to \cite{chang2025liftingwindingnumberprecise}. We represent a field $f : \Omega \rightarrow \mathbb{R}$, \emph{nonsmooth} across $\Gamma \subset \Omega$ as the restriction of an \emph{everywhere smooth} field $\Tilde{f} : \Omega \times \mathbb{R} \rightarrow \mathbb{R}$:
\begin{align}
f(\bm{x}) = \Tilde{f}(\bm{x},H(\bm{x})) \ .
\end{align}
We first lift each point $ \bm{x} \in \Omega $ into a higher-dimensional space by constructing the graph of $ H $:
\begin{align}
\label{eq:lifting_framework}
    \mathcal{L}(\bm{x}) = (\bm{x}, H(\bm{x}))
\end{align}
We then define the function  $f$  as the restriction of a smooth field  $\Tilde{f}$ back to the original domain via $ \mathcal{L}$:
\begin{align}
\label{eq:restriction}
    f(\bm{x}) = \Tilde{f}(\mathcal{L}(\bm{x})) \ ,\quad \bm{x} \in \Omega \ .
\end{align}

Notably, the domain of $\Tilde{f}$ is defined over the extrusion of the domain $\Omega$ along a new dimension, the lifting dimension. $f(\bm{x})$ is then discretized with a neural field with weights $\theta$, and we denote this by adding the subscript: $\tilde{f}_\theta$.

This approach sidesteps the challenge of directly learning discontinuities by representing the target function as smooth in the lifted domain, aligning with the inherent smoothness prior of neural representations. To achieve this, we design the function along the lifting dimension, $H(\bm{x})$, to be a $C^0$ function, which is continuous in value but with a discontinuous gradient across the designated interface $\Gamma$. Importantly, $H$ is given in closed form and does not involve any learned parameters.

\begin{figure}
\centering
\includegraphics[width=\linewidth]{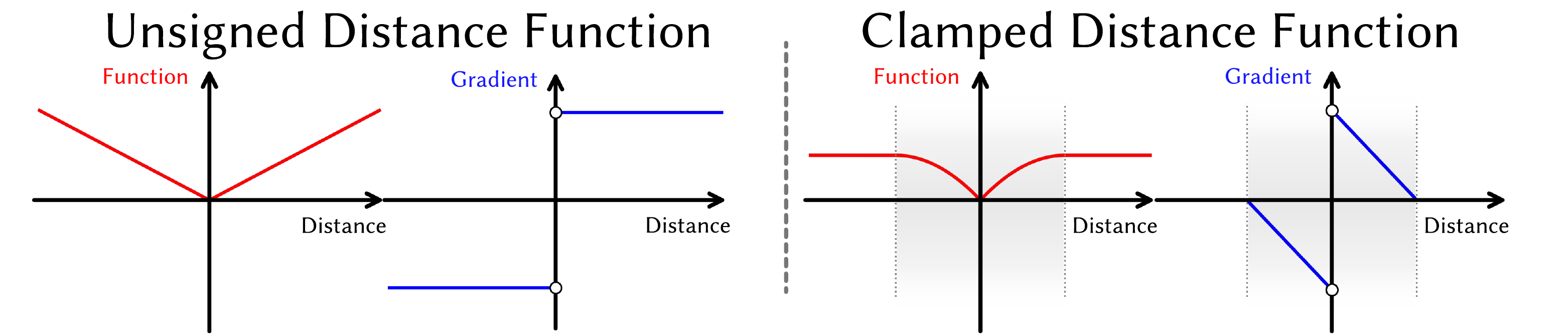}
\caption{We visualize the smoothly clamped distance function and its gradient. The function flattens beyond a threshold distance $s$, while preserving gradient discontinuities at the interface (where the distance is zero).}
\label{fig:didactic_smoothing}
\end{figure}

\paragraph{Choice of $H(\bm{x})$} 
We introduce a novel lifting strategy that departs from previous work on function-value discontinuities~\cite{chang2025liftingwindingnumberprecise}. Our goal is to represent a continuous function with discontinuous gradients, a property critical for capturing material interfaces and creases. We define a new lifting function \( H(\bm{x}) \) that is given in closed form and requires no learned parameters. This function is explicitly constructed from the interface geometry and designed to produce gradient discontinuities aligned with the interface normal.
%
We begin by computing the unsigned distance from the input coordinates to the interface:
\begin{align*}
D(\bm{x}) = \min_{\bm{p} \in \Gamma}\|\bm{x} - \bm{p} \|_2,
\end{align*}
where $\bm{p} \in \mathbb{R}^d$ is a point on the interface.
After discretization, the interface $\Gamma$ is represented by an explicit mesh $\mathcal{M} = \{V, E\}$. For a given point, we compute the closest distance to the nearest point on this mesh. We denote the distance function to the discrete interface $\mathcal{M}$ as $D_{\mathcal{M}}$.

Unsigned distance fields exhibit gradient discontinuities across the mesh due to their absolute-value-like behavior. While this may seem sufficient at first glance, the formulation has significant limitations. First, computing the unsigned distance field is inherently global: even for query points $\bm{x}$ located far from the interface, the computation may still involve traversing many primitives resulting in unnecessary overhead. Second, unsigned distance fields are known to produce singularities near the medial axis.
Although prior work~\cite{Madan:2022:smoothdists} addresses this by applying a softmin across primitives, the resulting formulation remains global. 

Since the key property we seek from the distance field is its absolute-value-like behavior near the interface, where gradient discontinuities occur, we do not need to compute exact distance values far from the interface; it is sufficient for the field to remain smooth in those regions. To this end, we adopt a smoothly clipped distance field that preserves the discontinuity near the interface while localizing computation within a specified threshold. This formulation also helps alleviate singularities caused by the medial axis by nullifying the gradient beyond the threshold region.

As shown in Figure~\ref{fig:didactic_smoothing}, the final height function is constructed by smoothly clamping the distance function: 
\begin{align*}
H(\bm{x}) = \|D_{\mathcal{M}}(\bm{x}) \|_{\text{SC}}
\end{align*}
where  $\|\cdot \|_{\text{SC}}$ is the smooth clamping function \cite{Chen:2023:LocalDeformation}:
\begin{align*}
\left\| {D}(\bm{x}) \right\|_{\text{SC}} = 
\begin{cases}
\left\| {D}(\bm{x}) \right\|_2 - \frac{1}{2s} \left\| {D}(\bm{x}) \right\|_2^2 ,& \text{if }  {D(\bm{x})} < s, \\
\frac{1}{2}s , & \text{if } D(\bm{x}) \geq s.
\end{cases}
\end{align*}

\begin{figure}
\centering
\includegraphics[width=\linewidth]{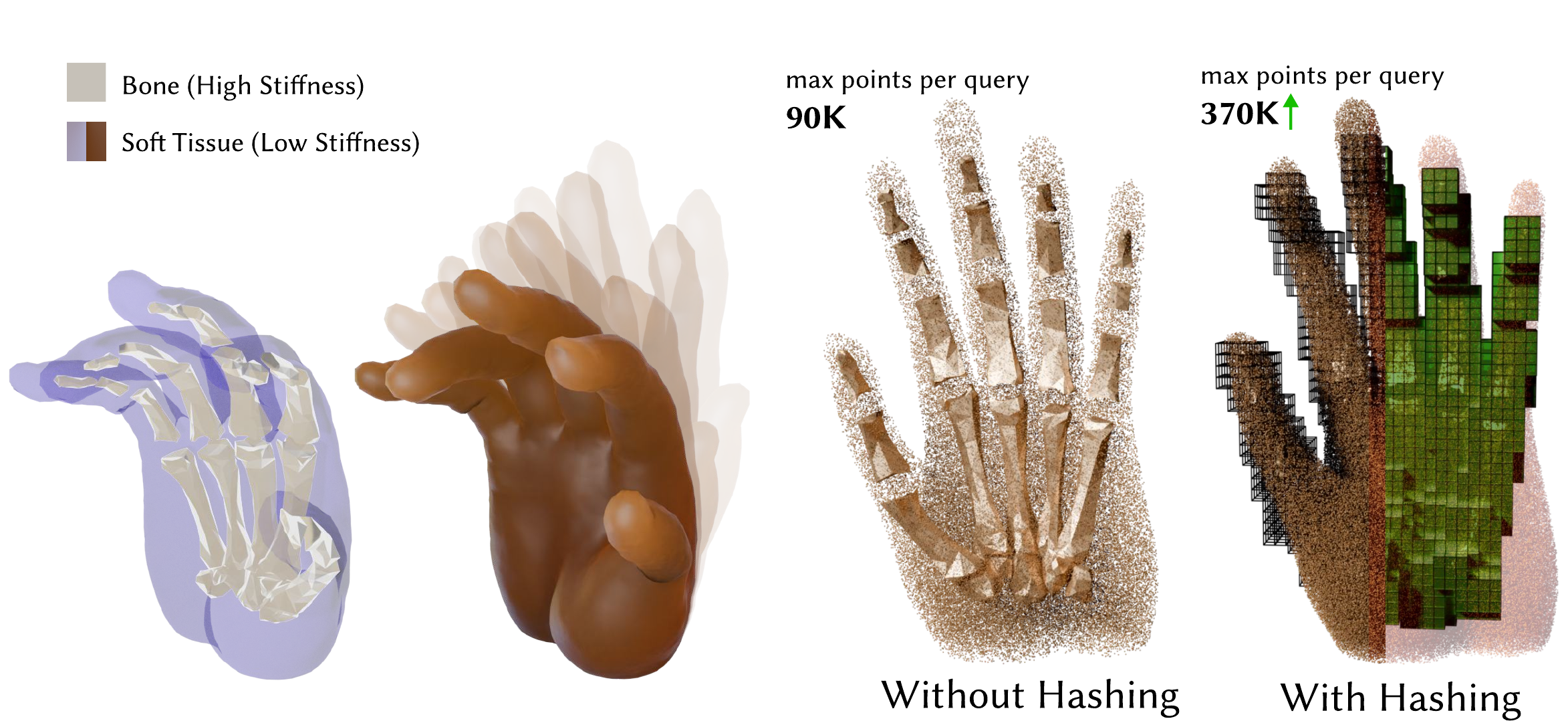}
\caption{We test our model on a complex scene of a hand with soft flesh and a stiff skeleton. This example demonstrates both reduced memory usage and speedup enabled by our spatial hash. The clamped distance function localizes queries by ignoring point pairs farther than $s$, making it well-suited for spatial hashing. The hash structure supports 4.1$\times$ more queries due to improved memory efficiency, and achieves a 3.6$\times$ speedup when tested with 90k query points (same as without hashing).}

\label{fig:hasing}
\end{figure}

This formulation has the advantage that only distances within a fixed threshold $ s $ need to be computed, making it well suited for acceleration structures such as spatial hashing. As shown in Figure~\ref{fig:hasing}, it requires 4.1 times less memory during queries and achieves a 3.6 $\times$ speedup compared to querying the same number of points without hashing.


\section{Adaptation to Reduced-space Simulation}



\subsection{Reduced-space Simulation}

We consider an elastic body on a reference domain $ \Omega \subset \mathbb{R}^d $, with deformation described by a displacement field $ \bm{u}(\bm{x}) $. In reduced-space simulation with skinning eigenmode subspace~\cite{benchekroun2023FastComplemDynamics, trusty:2023:subspacemfem}, this field is approximated using low-dimensional coordinates $ \bm{z} \in \mathbb{R}^k $ and basis functions $ \bm{\phi}_j $:
\begin{align*}
\bm{u}(\bm{x}) = \sum_{j=1}^{k} \bm{z}_j \bm{\phi}_j(\bm{x}) \begin{bmatrix} \bm{x} \\ 1 \end{bmatrix},
\end{align*}
where $ \bm{z}_j \in \mathbb{R}^{d \times (d+1)} $ and $ \bm{\phi}_j : \Omega \rightarrow \mathbb{R} $. Following recent work on neural reduced models~\cite{Modi:2024:Simplicits, chang:2023:licrom}, we represent basis functions using a neural field $ \bm{\phi}(\bm{x}) = f(\bm{x}) $. To support varying geometry or material properties, we condition on some 
parameter $\alpha$ (i.e., $\bm{\phi}^\alpha$), which corresponds to some (hand-built or learned) parameterization of the geometry (e.g., shape, material interfaces) or scenario (e.g., material stiffness, boundary conditions).




At each time step $ t $, the reduced coordinate $ \bm{z}_{t+1} $ is updated by minimizing the following energy:
\begin{align}
\label{eq:time_stepping}
\bm{z}_{t+1} = \argmin_{\bm{z}} \frac{1}{2} \left\| \bm{z} - 2\bm{z}_t + \bm{z}_{t-1} \right\|^2 + h^2 \int_{\Omega} \Psi\left( \bm{u}(\alpha, \bm{x}, \bm{z}) \right) \, d\bm{x},
\end{align}
where $ h $ is the time step size. The first term models inertia; the second is the elastic energy.

The energy $ \Psi $ depends on the spatial gradient $ \partial \bm{\phi}^{\alpha}_{\bm{x}} / \partial \bm{x} $, computed via automatic differentiation. The deformation gradient $ \mathbf{F} $ is formed as a linear combination of basis gradients and reduced coordinates. We approximate the integral using stochastic cubature with uniform sampling and optimize using gradient descent.

\subsection{Training on Varying Shapes and Material Stiffness}
\label{sec:training}
As mentioned before, we train a reduced model that adapts to a family of shapes and varying material properties. To capture these variations, we introduce a parameter $\alpha$ that encodes shape and stiffness information.

When $\alpha$ represents different shapes with changing interfaces, the interface $\Gamma^{\alpha}$ and its explicit representation $\mathcal{M}^{{\alpha}}$ both depend on $\alpha$. In our implementation, we treat $\mathcal{M}^{\alpha}$ as a nonlinear, user-defined function that aligns with the discontinuity interface. Accordingly, the height field becomes a function of $\alpha$:
\begin{align*}
H^{\alpha}(\bm{x}) = \|D_{\mathcal{M}^{{\alpha}}}(\bm{x}) \|_{\text{SC}}
\end{align*}

As $\alpha$ controls the shape and material variations, the network weights are also conditioned on $\alpha$ to adapt to the resulting changes in the domain. Thus, the parameter-conditioned neural field is defined as
\begin{align*}
f^{\alpha}(\bm{x}) = \Tilde{f}_{\theta}^{\alpha}(\mathcal{L}^{\alpha}(\bm{x})) \quad \textrm{and} \quad \mathcal{L}^\alpha(\bm{x}) = (\bm{x}, H^\alpha(\bm{x})).
\end{align*}

We parameterize the basis functions for reduced-space simulation using a neural field, denoted as $ \bm{\phi}^{\alpha}(\bm{x}) = f^{\alpha}(\bm{x}) $. To train the basis, we uniformly sample a parameter $ \alpha $ in each epoch, representing a domain $ \Omega^{\alpha} $, and then draw spatial samples $ \bm{x} \in \Omega^{\alpha} $ uniformly. The neural network weights $ \theta $ are optimized by minimizing a loss function evaluated on these samples:
\begin{align*}
L = E[\bm{\phi}^{\alpha}_{\theta}].
\end{align*}
The exact formulation of $ E[\bm{\phi}] $ depends on the application and will be discussed in detail in Section~\ref{sec:hetero_loss} and Section~\ref{sec:crease_loss}.

\begin{figure}
\centering
\includegraphics[width=\linewidth]{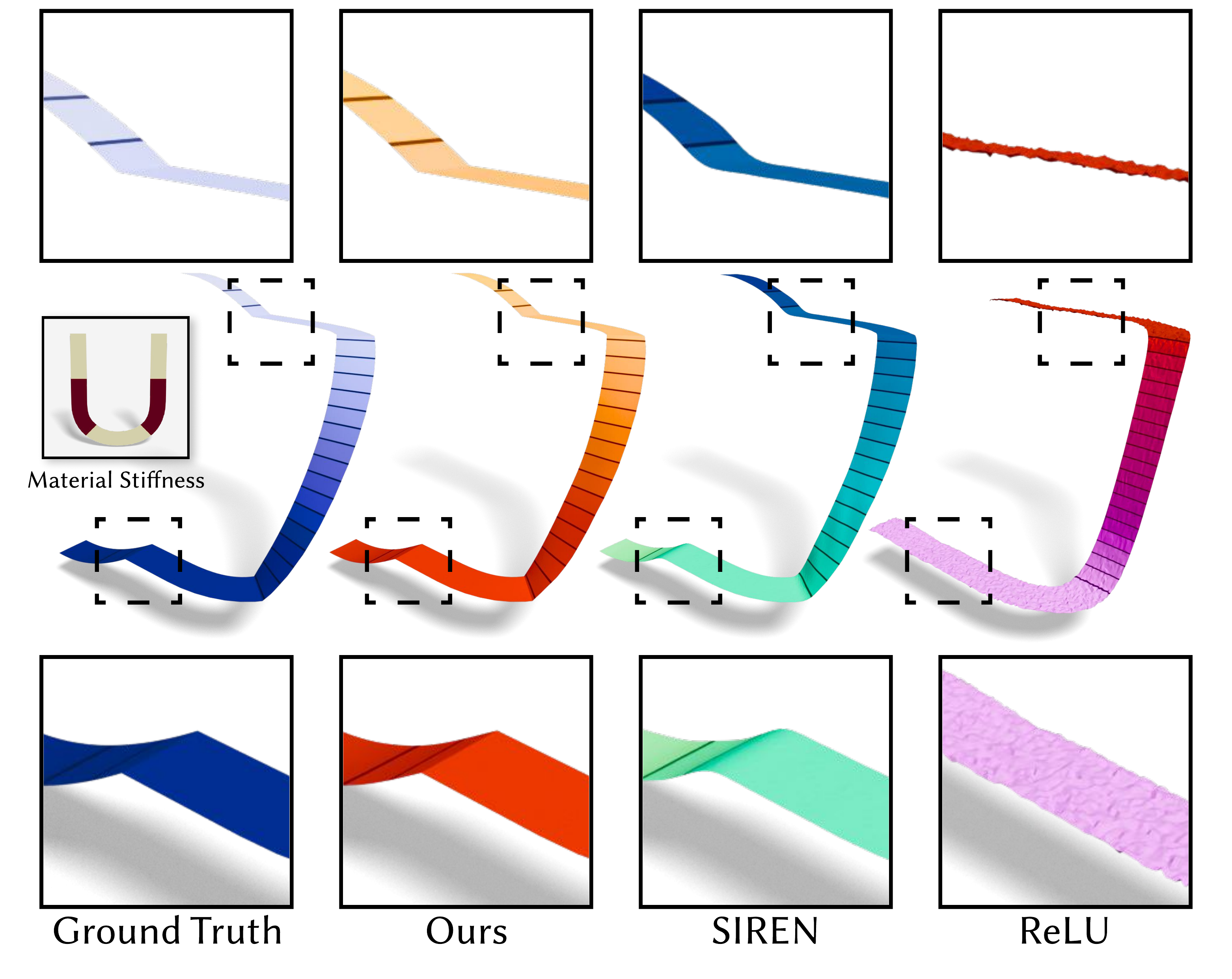}
\caption{We compared our method against basis functions produced by other neural field architectures on a heterogeneous 2D "U"-shaped domain. For visualization, the 2D shape is lifted along the Y-axis to represent the scalar basis function, with additional color coding to indicate its value. The SIREN MLP \cite{Sitzmann:2020:Implicit}, used in prior works \citep{chang2024neuralrepresentationshapedependentlaplacian, Modi:2024:Simplicits}, fails to capture sharp variations at the material interface. A ReLU-based neural field, which permits $C^0$ continuity, does not converge to the correct solution. In contrast, our method successfully captures the sharp gradient transitions across the interface.
}
\label{fig:heteroU}
\end{figure}

\subsection{Heterogeneous Elastodynamics}
\label{sec:hetero_loss}
Our method can be applied to compute basis functions for heterogeneous materials in a discretization-agnostic manner.

To construct the basis functions $\bm{\phi}$ for the skinning eigenmode subspace, we solve a generalized eigenvalue problem associated with the elastic energy Laplacian. In the case of co-rotational elasticity, this corresponds to a Laplacian operator weighted by the spatially varying material stiffness~\citep{benchekroun2023FastComplemDynamics}.

We extend the method of~\citet{chang2024neuralrepresentationshapedependentlaplacian} to heterogeneous materials by introducing a spatially varying weight function $w(\bm{x})$. Following their variational formulation, the eigenproblem is expressed as minimizing the weighted Dirichlet energy:

\begin{align} \label{eq:dirichlet} E_D[\bm{\phi}_i] = \frac{1}{2} \int_{\Omega} w(\bm{x}) |\nabla \bm{\phi}_i|^2  d\Omega, \end{align}
subject to the unit norm constraint $\bm{\phi}_i \in \mathcal{U}$, where  \\ $\mathcal{U} = \{ f \in L^2(\Omega) \mid |f|_2 = 1 \}$, and the orthogonality condition $\bm{\phi}_i \in \text{span}\{\bm{\phi}_1, \ldots, \bm{\phi}_{i-1}\}^{\perp}$. While their method handles only homogeneous settings, it fails to capture sharp material transitions due to the smoothness of the neural field. Our formulation overcomes this limitation

\paragraph{Gradient discontinuities in heterogeneous materials}
In heterogeneous materials, the varying stiffness induces gradient discontinuities across interfaces $\Gamma$, resulting in a \emph{jump condition} on the normal derivative~\cite{Gilbarg:2001:Elliptic},
\begin{align} w_1 \frac{\partial u}{\partial n} = w_2 \frac{\partial u}{\partial n}, \end{align}
where $ \frac{\partial u}{\partial n} $ denotes the normal derivative, and $\mathbf{n}$ is the unit normal to $\Gamma$. Refer to the supplemental material for a derivation.

While this behavior is familiar in mesh-based PDE contexts, it remains underutilized in neural simulation frameworks. Our method is the first to represent these discontinuities in a generalizable neural field architecture, enabling basis functions that remain valid across parametric shape and material spaces.

As illustrated in Figure \ref{fig:heteroU} and Figure \ref{fig:snail}, directly optimizing Equation~\ref{eq:dirichlet} using a standard MLP leads to overly smoothed results that fail to capture the sharp discontinuities at material interfaces. In contrast, our formulation preserves these transitions accurately. Unlike traditional mesh-based FEM approaches, where basis computation is tightly coupled to a specific discretization, our method is discretization-agnostic. This enables the computation of basis functions across a family of shapes (Figure \ref{fig:robot}), offering improved generalization.

\subsection{Creasing}
\label{sec:crease_loss}
Our method can also compute basis functions for subspace simulations of creasing, where the displacement field has discontinuous derivatives due to the crease. 

Unlike the previous section, which used a weighted Dirichlet energy, we compute the basis here by minimizing the Hessian energy \cite{Stein:2018:NBC:3191713.3186564}, subject to unit-norm and orthogonality constraints:
\begin{align} \label{eq:hessian} 
E_H[\bm{\phi}_i] \int_\Omega \|\nabla^2 \bm{\phi}_i\|_F^2 d\Omega \ ,
\end{align}
where $\|\cdot\|_F$ is the Frobenius norm and $\nabla^2 \bm{\phi} \in \mathbb{R}^{2 \times 2}$ denotes the second-order partial derivatives. The unit-norm and orthogonality constraints are implemented in the same fashion as in the previous section.
Note that this formulation does not impose explicit boundary conditions at the crease, the gradient change arises naturally from the neural field, which allows for discontinuities at the crease. Figure \ref{fig:crease_basis} shows the mode produced by our formulation, while traditional neural fields fail to capture such discontinuities due to their inherently smooth nature.

\section{Results}

\begin{table*}[ht]
\centering
\rowcolors{2}{white}{gray!20} 
\begin{tabularx}{\linewidth}{l |>{\centering\arraybackslash}p{1.2cm} |>{\centering\arraybackslash}X |>{\centering\arraybackslash}X |>{\centering\arraybackslash}X|>{\centering\arraybackslash}X|>{\centering\arraybackslash}X}
 & \textbf{\# basis} & \textbf{\# simulated vertices} & \textbf{\# vertices (interface)} & \textbf{\# elements (interface)} & \textbf{Basis building time (ms)} & \textbf{Time per step (ms)} \\
\textbf{Snail} (Figure~\ref{fig:snail})& 21 & 5.9k & 21 & 18 & 79.06 & 23.52 \\
\textbf{2D Robot}, $\alpha = 0$ (Figure~\ref{fig:robot}) & 7 & 19.0k & 98 & 93 & 37.42 & 19.71 \\
\textbf{2D Robot}, $\alpha = 0.33$ (Figure~\ref{fig:robot})& 7 & 25.5k & 98 & 93 & 41.71 & 23.03 \\
\textbf{2D Robot}, $\alpha = 0.67$ (Figure~\ref{fig:robot})& 7 & 48.7k & 98 & 93 & 68.60 & 37.91 \\
\textbf{2D Robot}, $\alpha = 1$ (Figure~\ref{fig:robot}) & 7 & 102.4k & 98 & 93 & 144.60 & 76.51 \\
\textbf{Bar }(Figure~\ref{fig:barbar}) & 4 & 40.0k & 44 & 40 & 86.28 & 2.78 \\
\textbf{Robot finger} (Figure~\ref{fig:finger})&  4 & 8.6k & 105 & 160 & 24.81 & 4.19 \\
\textbf{Shoe} (Figure~\ref{fig:shoeshoe})& 4 & 30.0k & 1924 & 2598 & 226.52 & 2.17 \\
\textbf{3D Robot }(Figure~\ref{fig:teaser})&  11 & 23.7k & 174 & 326 & 370.94 & 5.16 \\
\textbf{Crease }(Figure~\ref{fig:crease_basis}) & 4 & 150.0k & 20 & 19 & 87.42 & 5.34 \\
\textbf{Crease and Cut }(Figure~\ref{fig:kirigami}) & 2 & 50.0k & 104 & 101 & 7.75 & 1.87 \\
\end{tabularx}
\caption{We collected simulation timing data for the examples presented. Our method exhibits high performance, frequently achieving near real-time simulation speeds, even for cases involving a large number of simulated vertices. The reported basis building time reflects the duration required to perform network inference and construct the basis. While this step may introduce some overhead, it is only performed at the beginning of the simulation or when the design parameter $\alpha$ changes, and is therefore invoked infrequently.}
\label{tab:timing_table}
\end{table*}

All experiments are trained and tested on an NVIDIA RTX 4090 GPU. Our method is implemented in PyTorch and optimized using the Adam optimizer. The network architecture is a 5-layer, 128-channel SIREN MLP with positional encoding up to a maximum frequency of $ 2^5 $. The clamping factor is set to $ s = 1/8 $ for all 2D examples, $ s = 1/25 $ for the hand-with-skeleton example, and $ s = 1/16 $ for all other 3D cases.

\subsection{Heterogenous Materials}
\begin{figure}
\centering
\includegraphics[width=\linewidth]{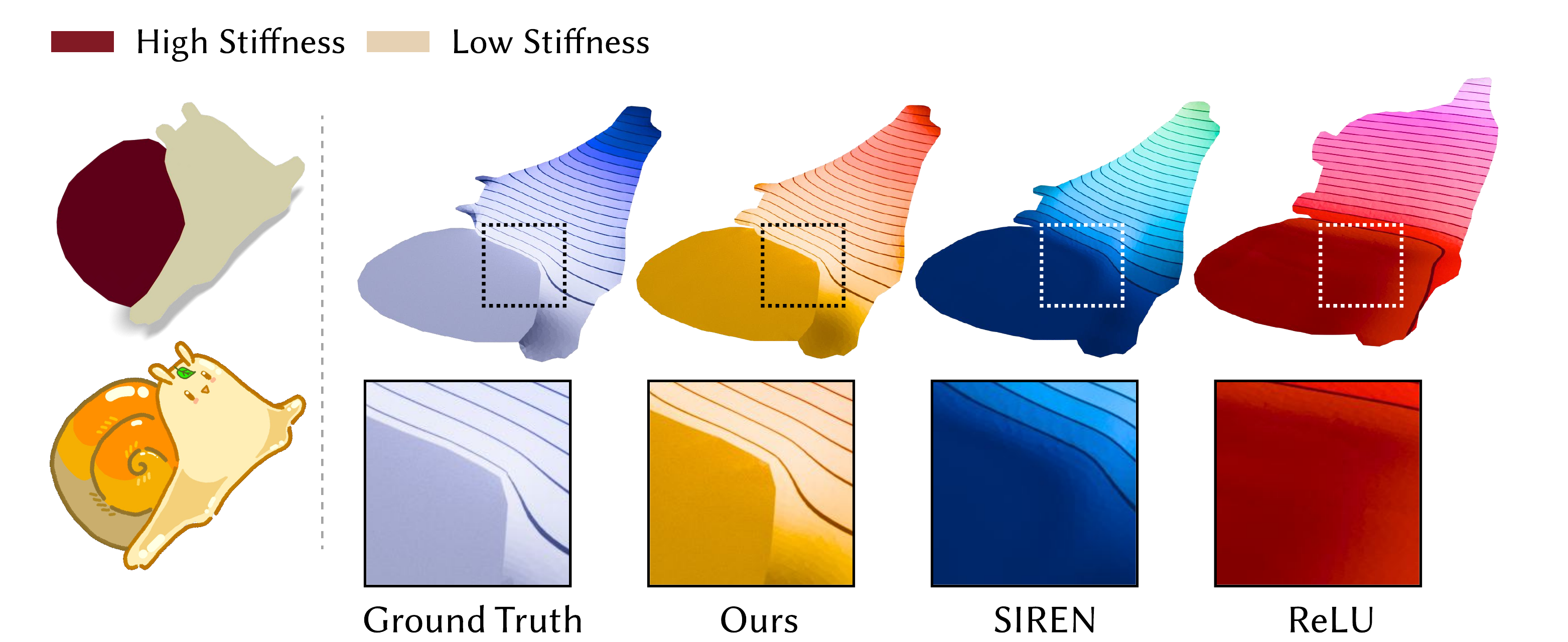}
\caption{We did another comparison on a snail shape where the shell is 100$\times$ stiffer than the body. SIREN and ReLU-based neural fields fail to capture the sharp gradient changes, while our method successfully recovers the gradient discontinuities.}
\label{fig:snail}
\end{figure}

\paragraph{Capturing Gradient Discontinuities} We demonstrate the ability of our basis to capture gradient discontinuities for a U shape (Figure \ref{fig:heteroU}) and a 2D snail (Figure \ref{fig:snail}). Our method is the only one to capture sharp gradient discontinuities at material stiffness changes, with results comparable to FEM, while being discretization agnostic—a point elaborated in the next paragraph. In contrast, a traditional neural field implemented using SIREN fails to capture the gradient discontinuity, exhibiting smoothing artifacts at the interface. We also implemented a neural field with a $C^0$ continuous activation function (ReLU). The hope was that such an activation function would allow the neural field to represent a $C^0$ continuous basis function. However, since the gradient discontinuity of ReLU cannot be aligned with the gradient discontinuity in the object, it fails to capture the gradient discontinuity at the stiffness interface. 

We can further expand our model to capture complex gradient discontinuities, as shown in Figure~\ref{fig:hasing}, which depicts a hand with soft flesh and a stiff skeleton simulated to curl up naturally.

\begin{figure}
\centering
\includegraphics[width=\linewidth]{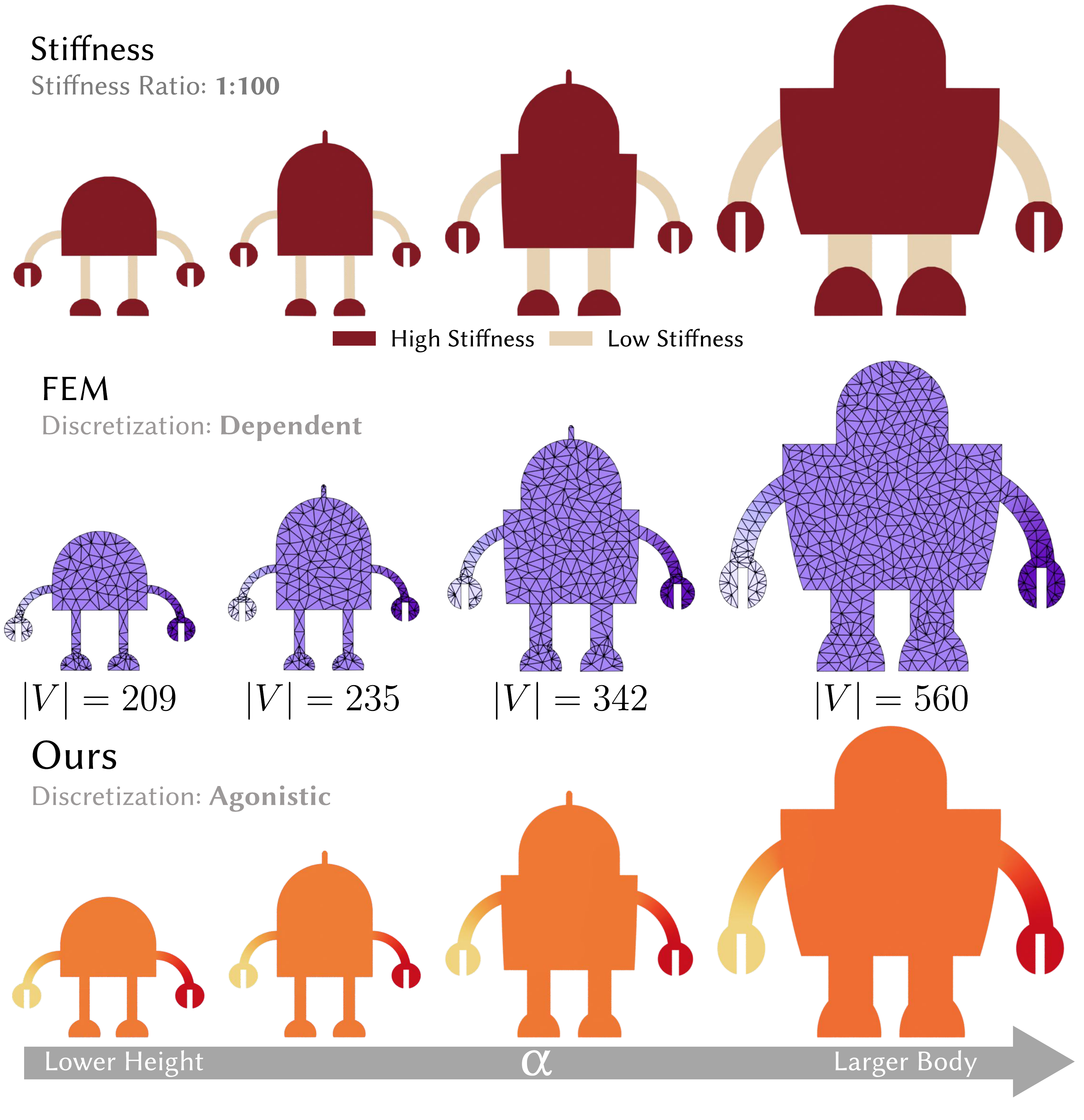}
\caption{Our model generalizes across heterogeneous domains, which is challenging for traditional methods based on eigenanalysis of discrete operators. We construct a robot shape space with varying material stiffness (first row) and apply constrained Delaunay triangulation \cite{Richard:2005:triangle}, resulting in meshes with 209–560 vertices. This variation hinders mesh-based methods from representing a consistent basis (second row), while our model captures basis functions across all shapes with a single network (third row). Scalar basis functions are visualized using color in the second and third rows.
}
\label{fig:robot}
\end{figure}

\paragraph{Discretization-Agnostic Representation} 
Figure~\ref{fig:robot} demonstrates the discretization-agnostic nature of our method. We train a neural field over a shape space of 2D robots with heterogeneous material distributions—specifically, softer arms and legs compared to the torso. While traditional FEM can capture gradient discontinuities, it requires explicit meshing of each domain. For shape families, maintaining consistent discretizations across all instances is often impractical or impossible, making it difficult to construct a shared basis across the space. As shown in the second row of Figure~\ref{fig:robot}, the number of vertices and faces varies significantly across shapes, preventing the reuse of a common reduced model.

In contrast, our method is independent of mesh discretization. A single trained neural field generalizes across the entire shape space, enabling basis inference on arbitrary geometries without remeshing, while maintaining accuracy comparable to FEM. This flexibility supports dynamic shape morphing during simulation without recomputation, as shown in the supplementary video. 

\begin{figure*}
\centering
\includegraphics[width=\linewidth]{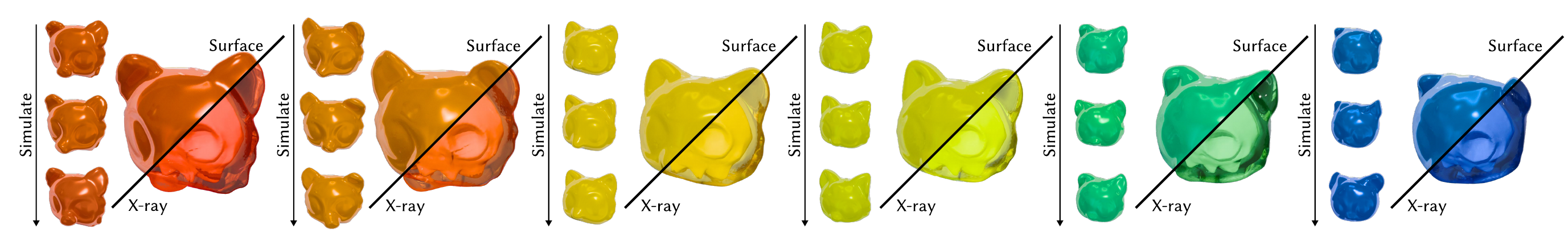}
\caption{Simulation of a parameterized shape family morphing from a fox to a bear, performing nodding and shaking motions. Each shape includes a skull that is $100 \times$ stiffer than the surrounding soft tissue, shown in the X-ray view. Our neural basis adapts across the family: soft regions (ears, nose) undergo large deformations, while the stiff skull remains rigid.
}
\label{fig:skull}
\end{figure*}

In Figure~\ref{fig:teaser}, we show a continuously evolving family of robots performing a dancing motion. These robots are sampled from a 3D shape space consisting of soft limbs and a stiff body. We train a single model to represent basis functions across the entire space. As a result, we can simulate the dancing motion with smoothly changing robot shapes, without remeshing or restarting the simulation.
Figure~\ref{fig:skull} shows another example: a family of cartoon animals with rigid skulls and soft tissue, parameterized by a variable \( \alpha \). The skull is 100 times stiffer than the surrounding head. As the animal nods and shakes its head, it morphs continuously from a fox to a bear. The neural basis adapts seamlessly: low-stiffness regions like the ears and nose exhibit large deformations, while the stiff skull remains relatively rigid.

\begin{figure}
\centering
\includegraphics[width=\linewidth]{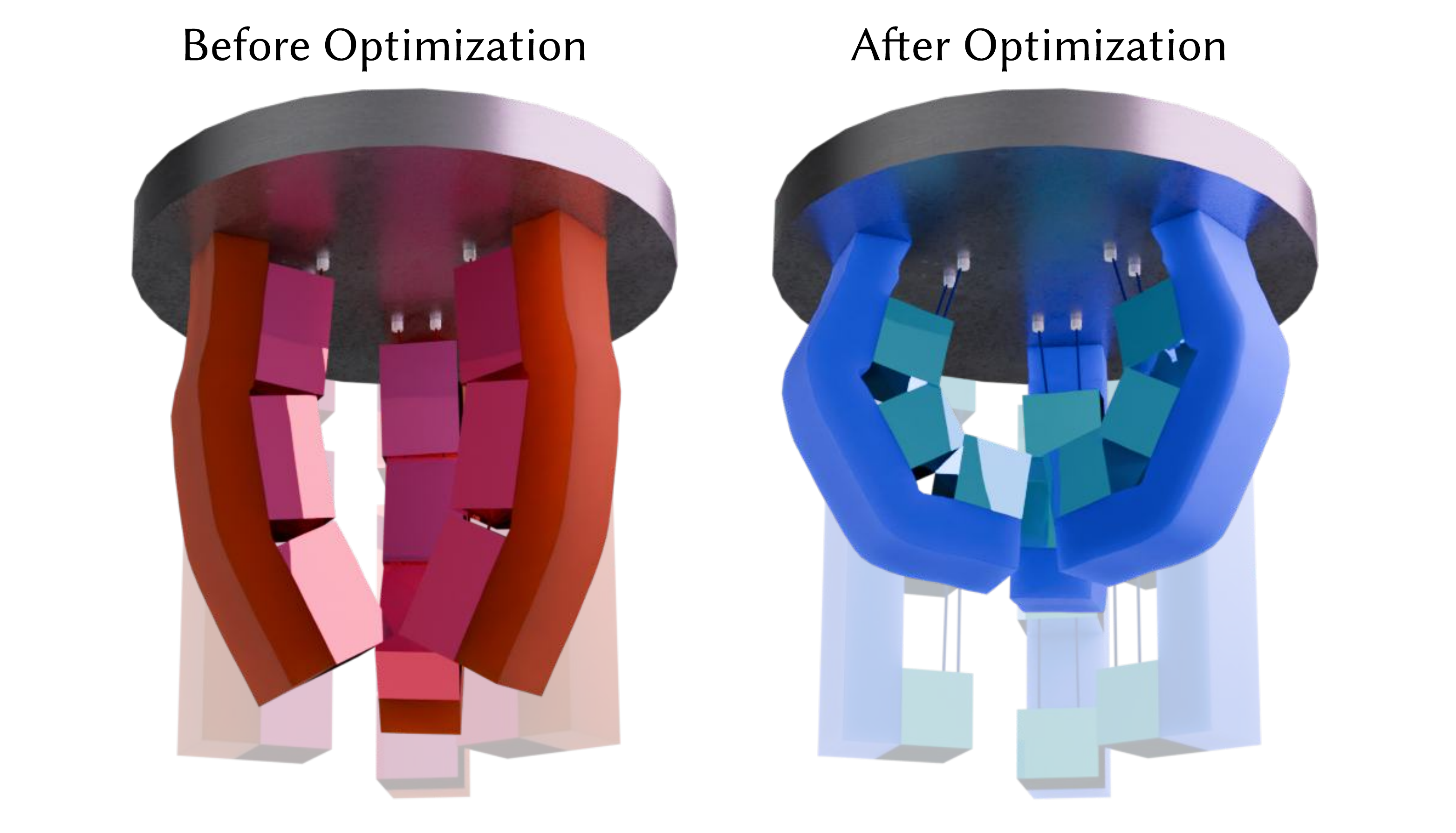}
\caption{Our model is differentiable with respect to the shape parameter $\alpha$, enabling shape optimization in reduced space. We optimize the spacing between bars in the robot fingers. The optimized shape (right) exhibits greater deformation compared to the initial guess (left).}
\label{fig:finger}
\end{figure}

\paragraph{Differentiable Subspace Physics Simulation} Our representation is differentiable with respect to the shape parameter $\alpha$, making it suitable for shape optimization tasks. 
Figure~\ref{fig:finger} and the insect illustrates an example involving objects resembling robotic fingers used for grasping.
\begin{wrapfigure}{r}{0.3\columnwidth}
    \hspace{-15pt}
    \includegraphics[width=0.32\columnwidth]{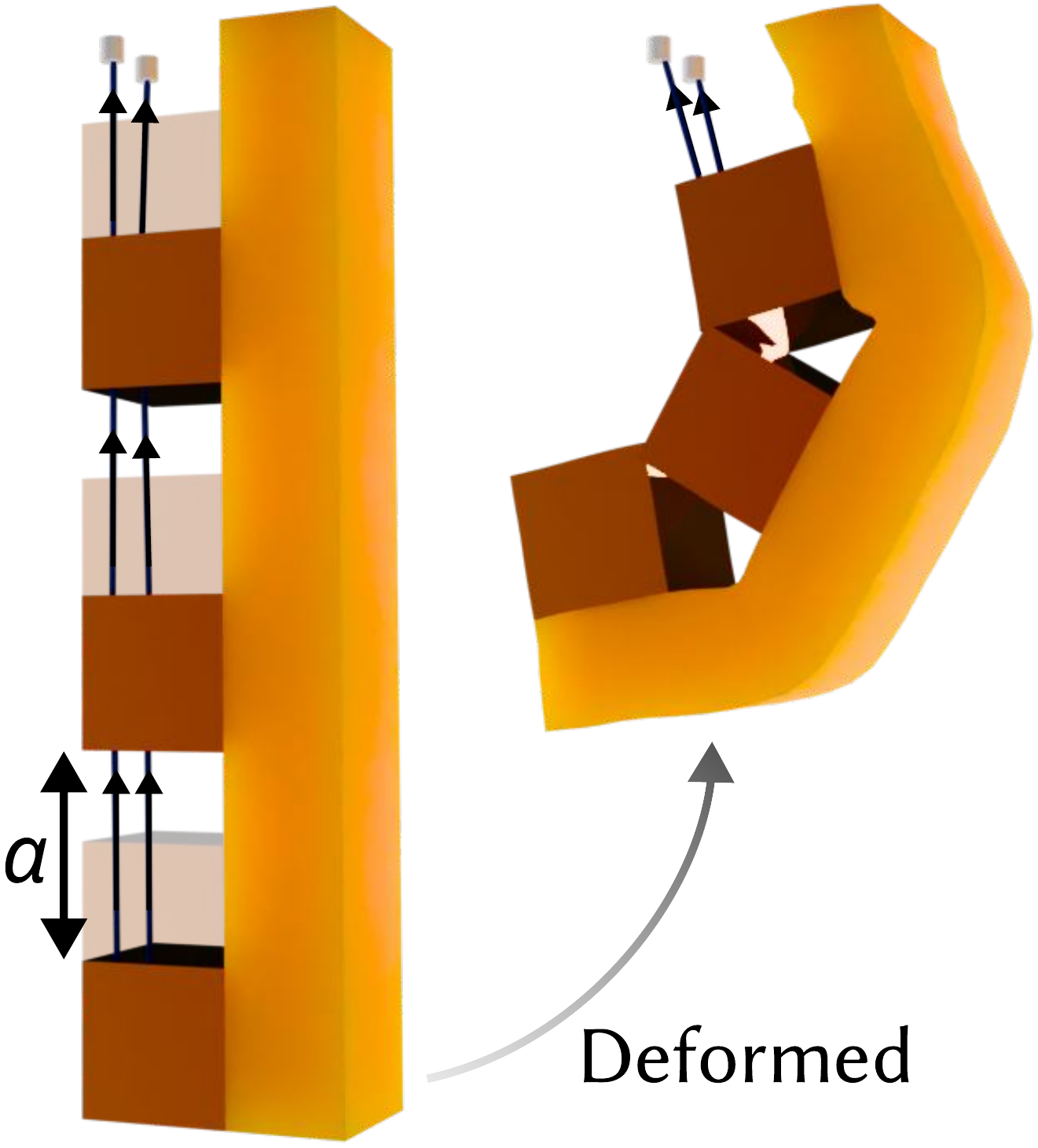}
    \vspace{-10pt}
\end{wrapfigure}
Each finger consists of a single long, soft cuboid bar, along with three smaller, stiffer blocks attached along one side. The grasping motion is controlled by adjusting the gap between the tops of the stiff blocks, as shown in the inset. Starting from an initial configuration, a more strongly bent finger shape is obtained by optimizing the shape code $\alpha$.

We perform forward simulation for 200 time steps, applying spring forces between the endpoints of each stiff segment to mimic strings stretching across different parts of the robot finger. We compute the mean final displacement along the $y$-axis, denoted as $\bm{u}_{\alpha}$, and determine the optimal shape parameter $\alpha^*$ by maximizing the displacement norm:
\begin{align*}
    \alpha^* = \argmax_{\alpha} \| \bm{u}_{\alpha} \|^2.
\end{align*}

As shown in Figure~\ref{fig:finger}, the optimized shape (right) bends significantly more than the initial guess (left), demonstrating the effectiveness of our method for shape optimization in reduced space.

\begin{figure}
\centering
\includegraphics[width=\linewidth]{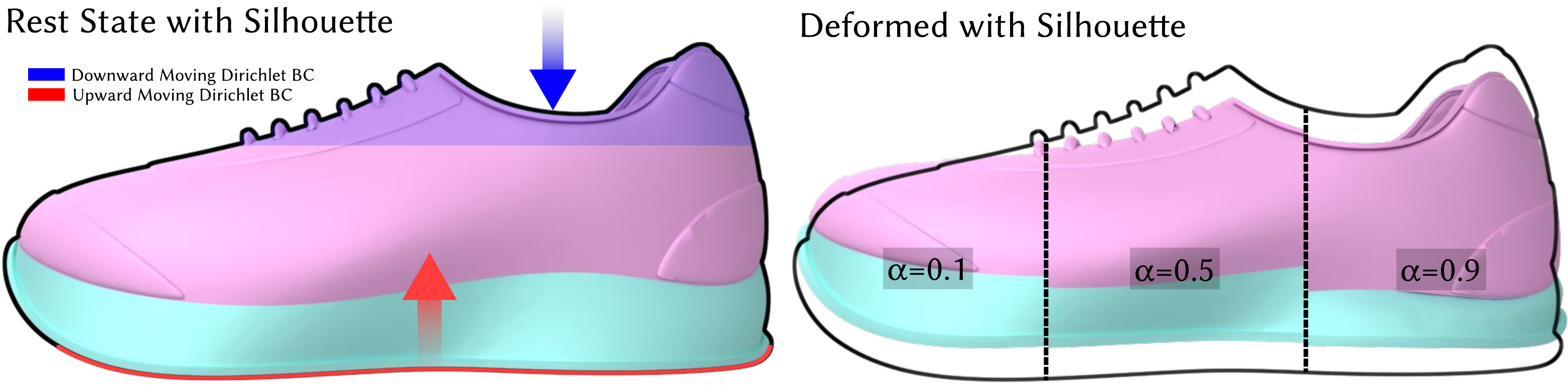}
\caption{We assign different stiffness values to the upper (pink) and sole (blue) regions of the shoe, controlled by a parameter $\alpha$. Higher $\alpha$ values increase the stiffness of the upper relative to the sole. When compressed from both above and below, the sole exhibits greater deformation as $\alpha$ increases. }
\label{fig:shoeshoe}
\end{figure}

\paragraph{Generalization}
Our framework generalizes not only across shape families but also over spatially varying material layouts, including parametric changes in stiffness. In Figure~\ref{fig:shoeshoe}, we vary the stiffness ratio between the upper part of the shoe (foot-facing) and the sole (ground-facing), distinguished visually by color. A material parameter $\alpha$ controls this ratio: larger $\alpha$ values increase the stiffness of the upper relative to the sole. We compress the shoe from both above and below and observe that as $\alpha$ increases, the sole becomes increasingly deformable, leading to greater compression in the lower region.

\begin{figure}
\centering
\includegraphics[width=\linewidth]{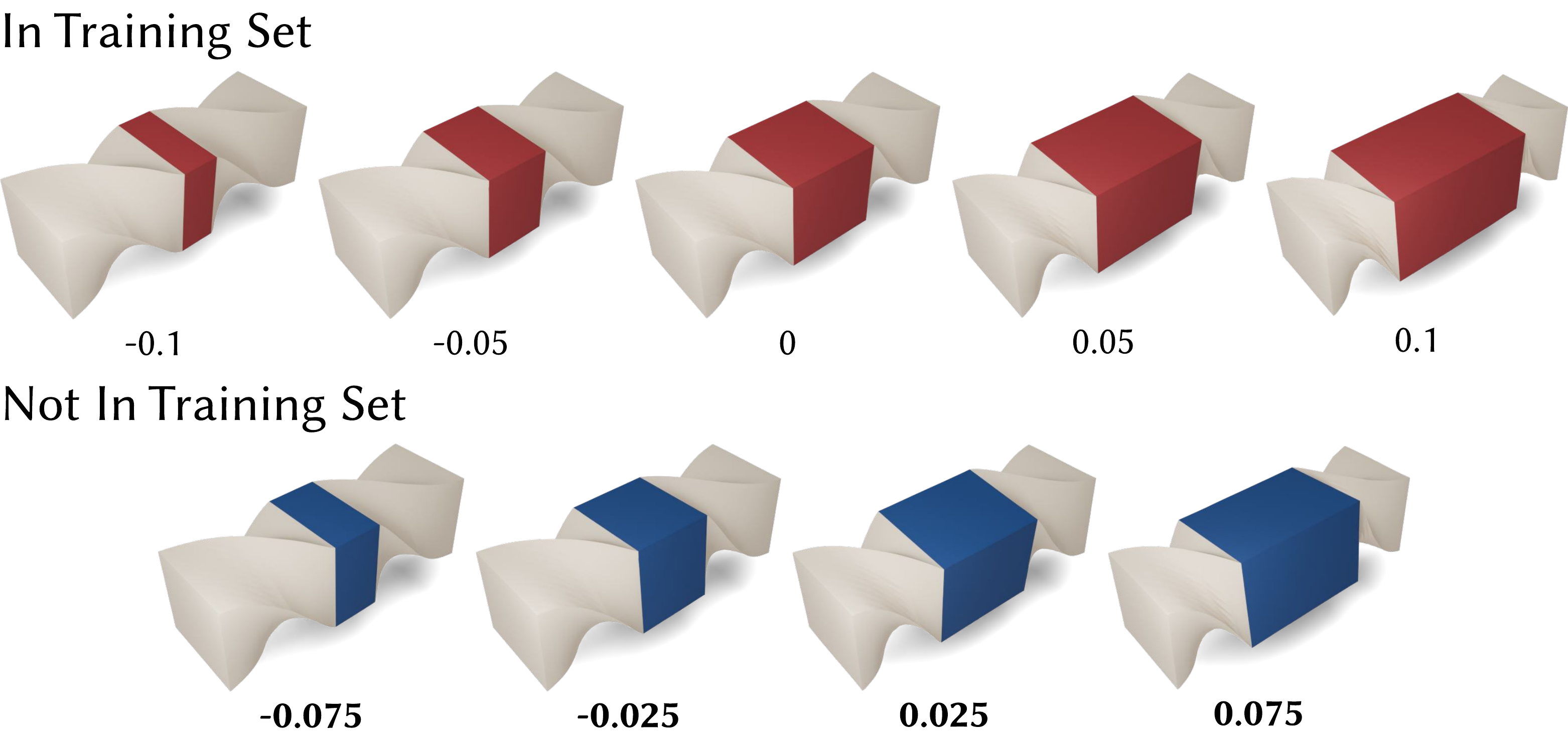}
\caption{Our method generalizes to unseen material configurations. In this cuboid bar example, the model was trained on only five distinct material layouts, each parameterized by $\alpha$. Despite this limited supervision, the simulation accurately handles both seen and unseen configurations, producing sharp transitions at stiffness boundaries.}
\label{fig:barbar}
\end{figure}

To further demonstrate the ability of our method to generalize across unseen material configurations, we evaluate it on a second example involving a bar with a cuboid rest state (Figure~\ref{fig:barbar}). Here, the basis for simulation was trained using only five different stiffness distributions, each parameterized by a material variable $\alpha$. We then performed twist simulations on bars with both seen and unseen $\alpha$ values. Our method not only delivers excellent simulation performance on examples from the training set, but also generalizes effectively to bars with stiffness distributions not encountered during training. It successfully creates sharp transitions at the boundaries between different materials while maintaining the structural integrity of the stiffer middle section.


\subsection{Creasing}

\begin{figure}
\centering
\includegraphics[width=\linewidth]{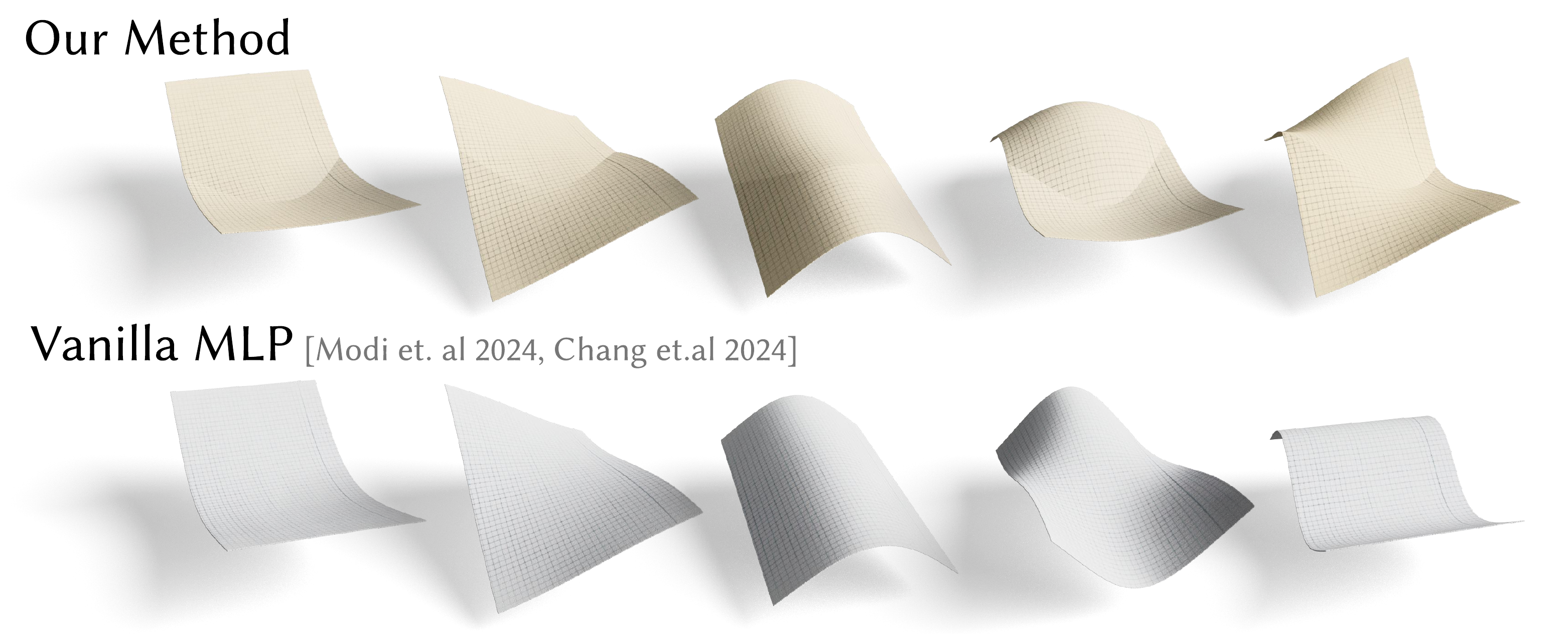}
\caption{Comparison of basis functions learned by our method (top row) and a standard SIREN MLP \cite{Modi:2024:Simplicits, chang2024neuralrepresentationshapedependentlaplacian} (bottom row) on a 2D domain with a crease. The scalar-valued basis functions are visualized by lifting along the $y$-axis. Our method successfully captures the sharp gradient discontinuity introduced by the crease, while the vanilla MLP fails to represent this non-smooth behavior due to its inherent continuity.}
\label{fig:crease_basis}
\end{figure}

\paragraph{Capturing Gradient Discontinuities} Our method captures the gradient discontinuity introduced by the crease. We train a 2D neural field using the loss defined in Equation~\ref{eq:hessian} and visualize the resulting basis functions by lifting the scalar-valued output along the $y$-axis. As shown in Figure~\ref{fig:crease_basis}, our approach (first row) successfully recovers the sharp edge along the crease, whereas a standard SIREN MLP (second row) fails to produce crease-aware basis functions due to its inherent continuity. Notably, the loss function in this example contains no explicit information about the discontinuity, as all spatial samples share the same stiffness. Thus, the gradient discontinuity is captured entirely by the network architecture, without relying on additional cues from the loss.

\paragraph{Interactive Designs of Crease Shapes} Since our boundary representation is explicit, the crease shape can be interactively edited during simulation. Using the basis functions from the previous example, we perform reduced simulations where the crease geometry is modified in real time. This allows the simulation to adapt immediately to user edits without requiring a restart, offering flexibility in editing the crease shape during runtime.

\paragraph{Modeling both Discontinuities in Function and Gradient} 
Since both our method and \cite{chang2025liftingwindingnumberprecise} adopt a lifting-based approach, we extend our framework to capture both function and gradient discontinuities, enabling simultaneous modeling of cuts and creases within a single model. This is achieved by incorporating their generalized winding number field as an additional dimension in our lifting function.

Specifically, we augment Equation~\ref{eq:lifting_framework} by adding an extra dimension for a second lifting function that captures function discontinuities, defined by the generalized winding number field from \cite{chang2025liftingwindingnumberprecise}:
\begin{align*}
    \mathcal{L}(\bm{x}) = (\bm{x}, H_{d}(\bm{x}), H_{gwn}(\bm{x})),
\end{align*}
where $H_{d}(\bm{x})$ is the smoothly clamped distance function from our method, and $H_{gwn}(\bm{x})$ is the generalized winding number field.

\begin{figure}
\centering
\includegraphics[width=1\linewidth]{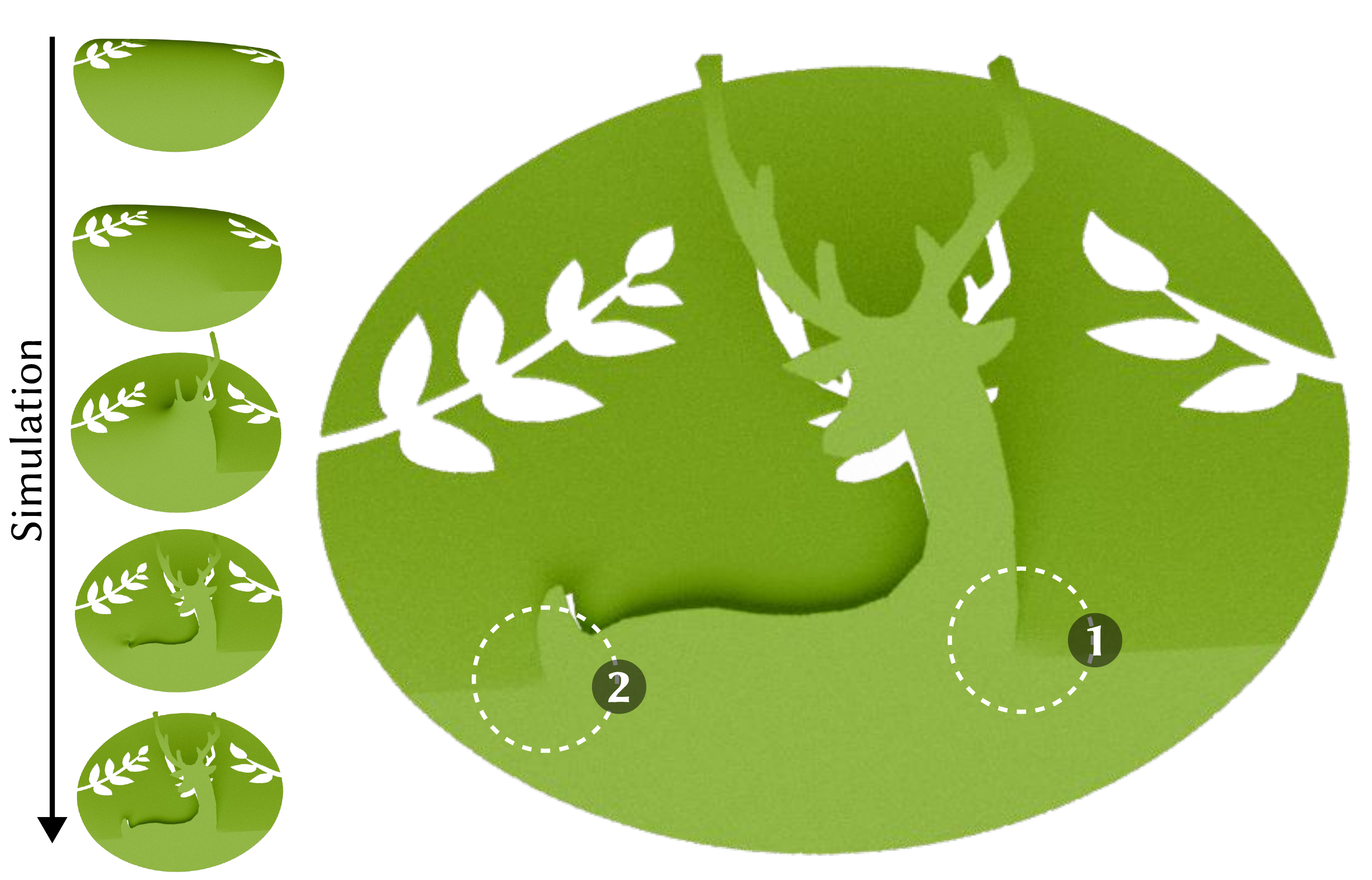}
\caption{Our method can be combined with \cite{chang2025liftingwindingnumberprecise}, enabling simulation of a paper-like material undergoing both progressive cutting and creasing. The sequence on the left shows the deformation evolving from smooth bending to sharp folds as creases and cuts develop. Region (1) highlights a transition from crease to cut, while region (2) shows the reverse: a transition from cut to crease.}
\label{fig:kirigami}
\end{figure}

As shown in Figure~\ref{fig:kirigami}, the paper deforms smoothly at the beginning of the simulation. As the crease develops, sharp turns emerge along it and, as the shape of the deer is progressively cut, the deformation becomes increasingly concentrated near the crease. This extension allows us to construct basis functions that are discontinuous in value, evident in the cuts of the deformed shape, while also exhibiting gradient discontinuities, such as those along the creases. This example demonstrates the flexibility and extensibility of our framework.

\section{Discussions and Future Work}
In this work, we proposed a novel neural field representation for capturing spatial gradient discontinuities and adapted it to reduced-space simulation of heterogeneous materials and creases. This enables several new applications, including shape optimization for reduced-order models of heterogeneous materials and simulations where cuts and creases evolve at runtime.

\begin{figure}
\centering
\includegraphics[width=1\linewidth]{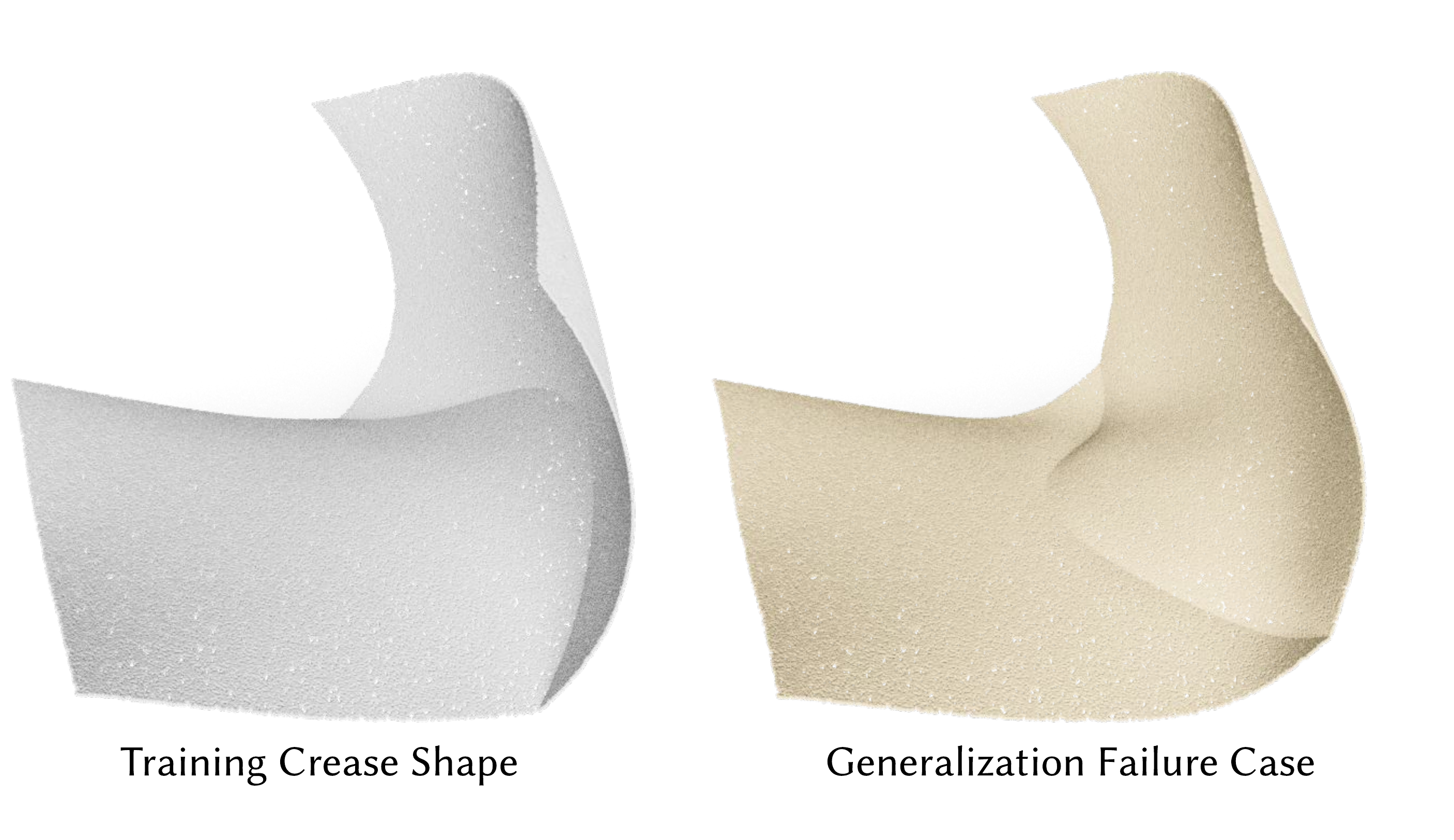}
\caption{We demonstrate a failure case (right) in generalization when the crease shape differs significantly from the one seen during training (left). While our method still captures the gradient discontinuity at the new crease location, the resulting behavior is less physically meaningful.}
\label{fig:failure}
\end{figure}

As with all methods based on neural networks, our approach has limitations in generalization when the test data distribution deviates significantly from the training set. 
In particular, when the test shape differs substantially from those seen during training, our method can still capture the visual appearance of the displacement field, including gradient discontinuities, but the resulting basis functions may be less physically meaningful. As shown in Figure~\ref{fig:failure}, when the test crease moves far from the training distribution, the displacement field remains discontinuous at the new crease location, but the paper no longer bends naturally along the crease. One possible way to alleviate this issue is to include more diverse training data by sampling additional values of $\alpha$.

A key property of our method is that the interface is explicitly represented and defined by the user. An interesting direction for future work is to automate this process or make it physics-driven, eliminating the need for manual alignment. This could potentially be achieved using differentiable techniques similar to those proposed in \cite{liu20242d}.

\bibliographystyle{ACM-Reference-Format}
\bibliography{main}

\clearpage

\begin{figure}
\centering
\includegraphics[width=1\linewidth]{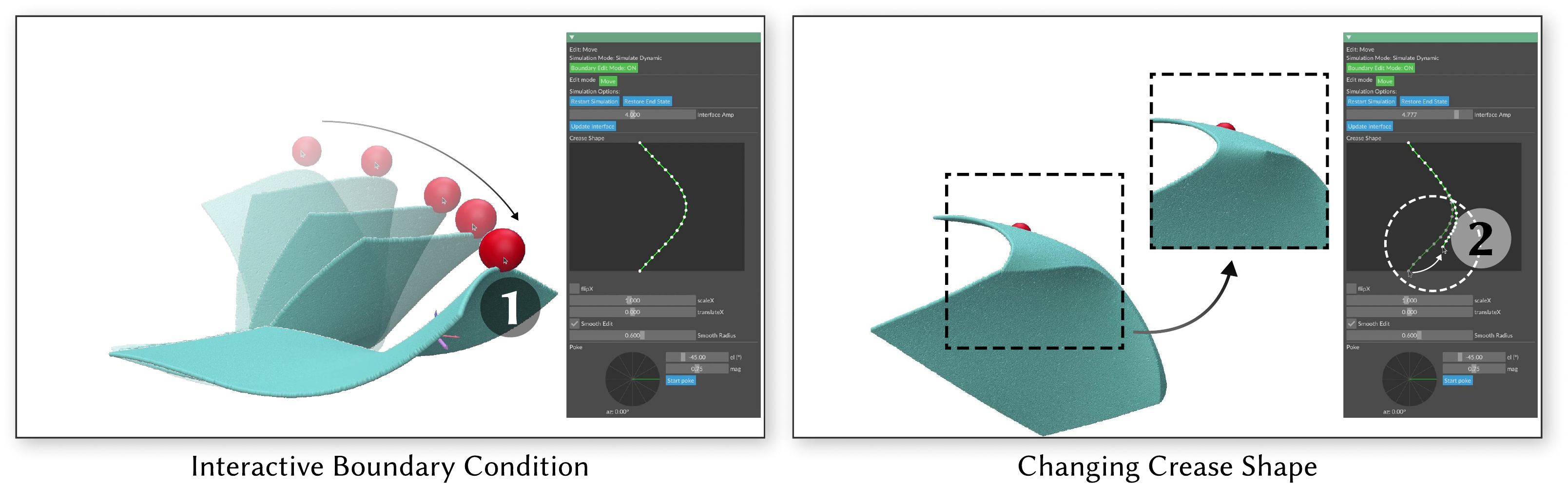}
\caption{Our method enables interactive editing of both crease geometry and boundary conditions. (1) highlights how dragging the red sphere changes the boundary condition, while (2) shows how the crease shape can be modified by editing the polyline directly in the interface.}
\label{fig:interactive}
\end{figure}

\clearpage

\end{document}